\documentstyle[12pt,prl,aps,psfig]{revtex}

\begin{document}
\draft
\title{Random Spin Signal in Magnetic Resonance Force Microscopy}
\author{G.P. Berman$^1$, V.N. Gorshkov$^{1,2}$, and V.I. Tsifrinovich$^3$}
\address{$^1$ Theoretical Division, Los Alamos National
Laboratory, 
Los Alamos, New Mexico 87545}
\address{$^2$ Department of Physics, Clarkson University, Potsdam, NY 13699}
\address{$^3$ IDS Department, Polytechnic University, Brooklyn, NY 11201}
\maketitle
\vspace{5mm}
\begin{abstract}
We study a random magnetic resonance force microscopy (MRFM) signal caused by the thermal vibrations of high frequency cantilever modes in the oscillating cantilever-driven adiabatic reversals (OSCAR) technique. We show that the regular MRFM signal with a characteristic decay time, $\tau_m$, is followed by a
  non-dissipative random signal with a characteristic time $\tau_r$. We present the estimates for the values of $\tau_m$ and $\tau_r$. We argue that this random MRFM signal can be used for spin detection. 
It has a ``signature'' of a sharp peak in its Fourier spectrum. 
\end{abstract}
\section{Introduction}
Magnetic resonance force microscopy (MRFM) is a powerful combination of magnetic resonance and atomic force microscopy which promises an atomic resolution for the magnetic resonance studies \cite{1,2}. When the number of detecting spins is small enough the fluctuating magnetic moment of a spin system becomes comparable with the maximum possible magnetic moment of the system. In this case, one expects to observe a random MRFM signal which follows a regular MRFM signal. This paper is devoted to the first theoretical analysis of this random MRFM signal. We consider the random signal caused by the high frequency thermal cantilever vibrations. 

Theoretical analysis for the initial (regular) MRFM signal in the OSCAR technique was performed in \cite{3,4}. In \cite{3} the Bloch type equations were derived to describe the motion of the local magnetization in the ``resonance slice'' of a sample. In these equations the relaxation terms are associated with the fluctuating magnetic field caused by the thermal vibrations of the cantilever. The approach used in \cite{3} implies no spatial correlations for the fluctuating magnetic field. As a result, the local magnetization in the resonant slice monotonically decays. In \cite{4} we have developed a computational model for simulations the MRFM signal in the OSCAR technique. Numerical simulations allow us to study both regular and random OSCAR MRFM signals. In this paper we use computer simulations to analyze the random MRFM signal.

In Section 2, we briefly explain the OSCAR technique, in particular, the sign of the frequency shift of the cantilever vibrations. In Section 3, we describe our numerical model. In Section 4, we present the qualitative analysis of the spin motion under the action of the fluctuating magnetic field caused by the high frequency cantilever vibrations. In Section 5, we describe the results of our numerical simulations of the regular and random MRFM signals.

\section{System under consideration}

We consider the ``parallel setup'' used in the experiment 
\cite{2}. (See Fig. 1.) 
We study the oscillating cantilever-driven adiabatic reversals (OSCAR) technique which is currently the most promising MRFM technique for spin-based atomic scale resolution \cite{2,5}. In OSCAR a feed-back loop is used to maintain a constant amplitude of the cantilever vibrations. The cantilever vibrations, in combination with an external {\it rf} magnetic field,
cause cyclic adiabatic reversals of the magnetic moment in the tiny ``resonance slice'' of the sample. In turn, the magnetic reversals generate an oscillating magnetic force on the cantilever tip which is approximately proportional to the cantilever amplitude. This force changes the net restoring force and, consequently, changes the frequency of the cantilever vibrations which is to be detected.

 \begin{figure}[t]
 \centerline{\psfig{file=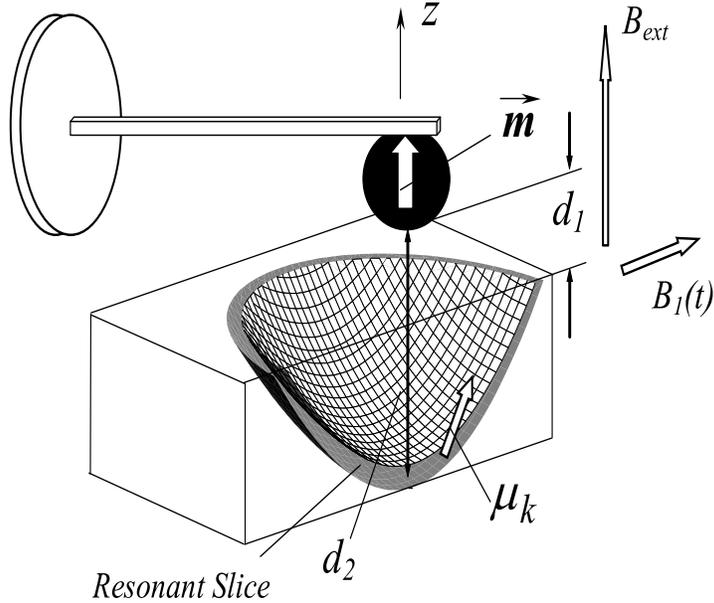,width=11cm,height=10cm,clip=}}
 \vspace{4mm}
 \caption{ The ``parallel setup'' used in the experiment 
  [2]. $B_{ext}$ and  $B_1(t)$ are the permanent and rotating 
 {\it rf} magnetic fields; $\vec m$ is the magnetic moment of the 
  ferromagnetic particle; $\vec\mu_k$ is the $k$-th magnetic moment 
  in the  resonant slice. }
 \label{fig:1}
 \end{figure}

The sign of the frequency shift can be easily understood from Fig. 1. Consider, for simplicity, only the magnetic moment $\vec{\mu}_k$ at the center of the resonant slice. When the $z$ coordinate of the ferromagnetic particle $z_c$ is positive, the distance between the particle and the center of the resonant slice is greater than the equilibrium distance. Consequently, the $z$ component of the magnetic field at the center of the resonant slice is less than the same quantity in the equilibrium position $z_c=0$. Then, in the rotating reference frame, the $z$ component of the effective field at the center of the resonant slice is negative. If the magnetic moment $\vec{\mu}_k$ points in the direction of the effective field (the ground state), then the $z$ component of the magnetic moment $\mu_z$ is also negative, unlike the magnetic moment of the cantilever tip. Because opposite magnetic moments repeal each other, the magnetic force on the ferromagnetic particle points in the positive $z$ direction, opposite to the cantilever spring force. As a result, the cantilever frequency decreases. Based on the same arguments, if the magnetic moment in the resonant slice is opposite to the direction of the effective field, its magnetic force causes the cantilever frequency to increase.

\section{The model for numerical simulations}

We describe the vibrations of the cantilever tip or, more precisely, the center of the ferromagnetic sphere, as the motion of a harmonic oscillator with a fundamental frequency $\omega_c$ and a quality factor Q:
$$
\ddot z_c+z_c+\dot z_c/Q=f(\tau).\eqno(1)
$$
In this equation, we use the dimensionless coordinate $z_c$ of the center of the ferromagnetic particle (in terms of the amplitude of the cantilever vibrations $z_0$), and the dimensionless time $\tau=\omega_ct$, $f(\tau)$ is the dimensionless magnetic force produced by the spins of the resonant slice on the ferromagnetic particle:
$$
f(\tau)=\sum_{k=1}^{\rm N}\eta_k\mu_{kz},\eqno(2)
$$
$$
\eta_k={{\mu_0}\over{4\pi}}{{3m\mu}\over{k_cz^5_0}}{{z_k(5{ z}_k^2-3{r}^2_k)}\over{{r}^7_k}},~\vec{r}_k=(x_k,y_k,z_k).
$$
Here N is the number of magnetic moments in the resonant slice, ${\vec \mu}_k$ is the $k$-th magnetic moment (in terms of its magnitude $\mu_k$), $k_c$ is the effective spring constant of the cantilever, $m$ is the magnetic moment of the ferromagnetic particle, $\vec r_k=\vec r_k(z_c)$ is the vector connecting the center of the vibrating ferromagnetic sphere with the position of the $k$-th magnetic moment.

We assume the number of spins N in the resonant slice is large enough, so that we can ignore quantum effects and consider classical equations of motion for the magnetic moments $\vec \mu_k$, in the rotating frame:
$$
\dot \mu_{kx}=-\Delta_k \mu_{ky},\eqno(3)
$$
$$
\dot \mu_{ky}=\Delta_k \mu_{kx}-\varepsilon\mu_{kz},
$$
$$
\dot \mu_{kz}=\varepsilon\mu_{ky}.
$$
Here
$$
\Delta_k=(\gamma B_{ext}-\omega)/\omega_c+{{\mu_0}\over{4\pi}}{{\gamma m}\over{\omega_cz^3_0}}{{3 z^2_k-{r}^2_k}\over{{ r}^5_k}},\eqno(4)
$$
$$
\varepsilon=\gamma B_1/\omega_c,
$$
$\gamma$ is the magnitude of the spin gyromagnetic ratio, $\omega$ is the frequency of the {\it rf} field. 

To simulate the thermal noise caused by the high frequency cantilever modes, we replace the vector $\vec r_k(z_c)$ in the expression for $\Delta_k$ by the vector $\vec r_k(z_c+\delta z_c)$, where 
$$
\delta z_c=\sum_n2a_n\cos(\Omega_n\tau+\Psi_n).\eqno(5)
$$
In Eq. (5), $a_n$ and $\Omega_n$ are the dimensionless thermal amplitude and the frequency of the $n$-th cantilever mode, and $\Psi_n$ is a random phase. The factor 2 in Eq. (5) appears because the amplitude of the cantilever tip for any mode $n$ is twice the amplitude of the mode \cite{6}. The thermal amplitude of a high frequency mode can be estimated from the equipartition theorem
$$
a_n=(z_0\Omega_n)^{-1}(k_BT/2k_c)^{1/2},\eqno(6)
$$
where $T$ is the cantilever temperature. In Eq. (6) we took into account the expression for the cantilever fundamental mode frequency  $\omega_c^2=4k_c/m_c$, where $m_c$ is the mass of the cantilever.
(See, for example, \cite{7}.) 

\section{Qualitative analysis and estimates}

Consider the motion of a single classical magnetic moment $\vec \mu_k$ in the center of the resonant slice, under the action of the fluctuating magnetic field produced by the ferromagnetic particle. Assume that the magnetic moment moves adiabatically, together with the effective field in the semi-plane (+z)-(+x)-(-z), in the rotating frame. When the polar angle of the vector $\vec\mu_k$ is not small, the weak random fluctuating field produced by the ferromagnetic particle has a component perpendicular to $\vec\mu_k$ which causes a deviation from the effective field. The resonance frequency $\omega_e$ in the rotating frame is 
$$
\omega_e=(\omega_z^2+\omega_R^2)^{1/2},\eqno(7)
$$
where $\omega_z=\omega_z(z_c)$ is associated with the $z$ component of the effective field which is proportional to the ferromagnetic particle coordinate $z_c$, and $\omega_R=\gamma B_1$ is the Rabi frequency associated with the {\it rf} field. The maximum value of $\omega_z$, $\omega_{zm}$, is much greater than $\omega_R$. In the process of adiabatic reversals the resonance frequency of the magnetic moment changes from $\omega_{zm}$ (near the (+z) axis) to $\omega_R$ (near the transverse plane), and back to $\omega_{zm}$ (near the (-z) axis). If a characteristic frequency of the fluctuating field falls in the region ($\omega_R,\omega_{zm}$) the fluctuating field causes a noticeable deviation of the magnetic moment from the effective field. We consider the fluctuating field generated by high frequency cantilever modes which cover the interval ($\omega_R,\omega_{zm}$). The thermal amplitudes of these modes are inversely proportional to their frequency. (See Eq. (6).) As an example, Fig. 2 demonstrates the thermal amplitudes of the cantilever modes in the interval ($\omega_R,5\omega_R$) for the data presented in experiment \cite{2}. 
One can see that among the high frequency modes which can be resonant to the magnetic moment, the greatest thermal amplitudes have the modes in the narrow region, approximately ($\omega_R,2\omega_R$). The geometrical factor also favors these modes because the magnetic moment $\vec\mu_k$ has the resonant frequency in the region ($\omega_R,2\omega_R$) when it is very close to the transverse plane where the fluctuating field is perpendicular to the magnetic moment $\vec\mu_k$.  
To estimate the action of the fluctuating field, we will consider only the modes in the narrow region ($\omega_R,2\omega_R$). We will assume that in this region of the resonant frequencies the magnetic moment experiences the fluctuating magnetic field produced by the thermal vibrations of the cantilever. To simplify our estimate, we assume that these vibrations have the constant amplitude $a_T$ and the random phase. We estimate the amplitude $a_T$ using Eq. (6) and substituting $\Omega_n\rightarrow\omega_R/\omega_c$:
$$
a_T=(\omega_c/\omega_R)(k_BT/2k_c)^{1/2}.\eqno(8)
$$
The corresponding thermal amplitude of the ferromagnetic particle on the cantilever tip is $2a_T$. When the magnetic moment passes the region of the resonant frequencies ($\omega_R,2\omega_R$), it experiences a deviation from the effective field. The amplitude of the fluctuating field is $Ga_T$ where $G$ is the magnetic field gradient, and factor 2 disappears as we consider the rotating component of the fluctuating field.  

The characteristic time between the ``phase jumps'' of the fluctuating field (the correlation time $\tau_0$) can be estimated as the reciprocal of the frequency region $\tau_0\approx 1/\omega_R$. Assuming that the angular deviation from the effective field is 
a diffusion process, we obtain 
$$
\overline{\Delta\Theta^2_1}=D\Delta t_1,\eqno(9)
$$
where $\overline{\Delta\Theta^2_1}$ is the characteristic deviation during a single reversal, $D$ is the diffusion coefficient, and $\Delta t_1$ is the time of passing the region ($\omega_R,2\omega_R$) during one reversal. The value of $D$ can be estimated as $\Delta\Theta^2_0/\tau_0$, where $\Delta\Theta_0$ is the characteristic deviation between two consecutive phase jumps, $\Delta\Theta_0=\gamma Ga_T\tau_0$. The time interval $\Delta t_1$ can be easily estimated from the equation $\omega_e=2\omega_R$. Substituting into Eq. (7)
$$
\omega_z=\gamma Gz_0\cos\omega_ct,\eqno(10)
$$
we obtain
$$
\Delta t_1=2\sqrt{3}(\gamma Gz_0)^{-1}(\omega_R/\omega_c).\eqno(11)
$$
(Note that the number of phase jumps during one reversal can be estimated to be $\Delta t_1/\tau_0=\omega_R\Delta t_1$. For example, for experiments \cite{2}, where $\omega_c/2\pi=21.4$ kHz, 
$G=1.4\times 10^5$ T/m, $z_0=28$ nm, $\omega_R/2\pi$ ranges between $4.17$ MHz and $8.34$ MHz, the number of jumps ranges between 25 and 100.) Combining Eqs. (9) and (11) we find
$$
\overline{\Delta\Theta^2_1}=(2\sqrt{3})(\gamma Ga^2_T/\omega_cz_0).\eqno(12)
$$

Now we can estimate how many reversals are needed to decrease twice the component of the magnetic moment along the effective magnetic field. To get the total deviation $\overline{\Delta\Theta^2}\approx 1$ we need $1/\overline{\Delta\Theta^2_1}$ reversals. The corresponding time $\tau_m$ is
$$
\tau_m=\pi/(\omega_c\overline{\Delta\Theta^2_1})=(\pi/2\sqrt{3})(z_0/\gamma Ga^2_T).\eqno(13)
$$
Substituting in Eq. (13) the expression for $a_T$ in (8), and omitting the numerical factor $\pi/\sqrt{3}$, we obtain finally the required estimate
$$
\tau_m\sim (k_cz_0/\gamma Gk_BT)(\omega_R/\omega_c)^2.\eqno(14)
$$

We shall discuss now this formula. The characteristic time of the deviation from the effective field increases with increasing Rabi frequency $\omega_R$, but decreases with increasing temperature. These dependences reflect the obvious properties of the thermal noise of the cantilever. The dependence $\tau_m$ on the amplitude of the cantilever vibrations $z_0$ is associated with the time $\Delta t_1$ of passing the resonant region ($\omega_R,2\omega_R$): the greater $z_0$ is, the smaller time $\Delta t_1$; thus the greater number of reversals is needed to provide a significant deviation of $\vec\mu_k$ from the effective field. The dependence $\tau_m$ on the magnetic field gradient $G$ appears as the result of two competing factors. On the one hand, when the gradient $G$ increases, the fluctuating magnetic field also increases. On the other hand, the time of passing the resonant region $\Delta t_1$ decreases when the gradient $G$ increases. 

Now consider the motion of the magnetic moment in the  ``tilted system'' connected to the effective field. In the ``titled system'' a single classical magnetic moment moves randomly between the direction of the effective field and the opposite direction. If the characteristic time of the deviation from the effective field is $\tau_m$, then after time $\sim 2\tau_m$ the MRFM signal changes from its maximum value $\Delta\omega_m$ to zero. After this, the MRFM signal may change randomly with the characteristic time $\tau_r\sim 8\tau_m$, sometimes approaching its maximum value $\Delta\omega_m$. (Correspondingly, a single quantum spin 1/2 is expected to jump randomly between two directions of the effective magnetic field with the characteristic time $\tau_r\sim 8\tau_m$. In the quantum case, the MRFM signal will take only two values $\pm\Delta\omega_m$.) For a large number of spins in the resonant slice, the quantum and classical systems have a similar behavior: one may expect first a ``regular decay'' of the MRFM signal with the characteristic time $\tau_m$, and then the random MRFM signal of a smaller amplitude with the characteristic time $\tau_r\sim 8\tau_m$. The amplitude of the random signal is expected to be smaller than the maximum value of the regular signal $\Delta\omega_m$ because spins in the resonant slice change their directions at different times. So, they may experience different fluctuating fields and have different components along the effective field.

Finally, if the number of spins in the resonant slice is very large, one can choose a small (but still macroscopic with many spins) volume of averaging $\Delta V$, where all spins experience the same effective field (the external field and the field produced by the ferromagnetic particle in the rotating frame). After this, one can introduce a continuous magnetization in the resonant slice 
$$
\vec M(\vec r)=\sum\vec\mu_k/\Delta V,
$$
where the sum is taken over the spins in the volume $\Delta V$ near the point $\vec r$. Note that this approach assumes that the effective field inside the volume of averaging is approximately uniform. So, the magnitude $M$ is constant. To describe the change of the magnitude $M$ due to the microscopic relaxation mechanisms (e.g. the spin-phonon interaction) one adds the relaxation terms in the equations of motion, e.g. Bloch type relaxation terms. In our case, the fluctuating field is produced by the ferromagnetic particle. The scale of the spatial inhomogeneity for the fluctuating field is the same as for the regular magnetic field produced by the ferromagnetic particle. Thus, we expect that the local magnetization magnitude $M(\vec r)$ will conserve until the microscopic relaxation mechanisms come into play. As a result, we may expect that the random MRFM signal will be observed even for a continuous magnetization.

\section{Numerical simulations}

In this section, we present the results of computer simulations of the random MRFM signal in the OSCAR technique. We solved Eqs. (1) and (3) with thermal noise (5). The magnetic moments were uniformly distributed in the resonant slice. The upper and the lower boundaries of the resonant slice have been determined from the equations $\Delta_k=0$ at $z_c=\pm 1$. Each time the cantilever passes the upper point, the computer automatically increases the value of $z_c$ to 1. This procedure corresponds to the action of a feed-back loop in OSCAR which maintains a constant amplitude of the cantilever oscillations. The main parameters of the problem were taken from the experiment \cite{2}
$$
B_{ext}=140 {\rm mT},~k_c=0.014 {\rm N/m},~\omega_c/2\pi=21.4 {\rm kHz},~z_0=28 {\rm nm},~Q=2\times 10^4,
$$
$$
m=1.5\times 10^{-12}{\rm J/T},~M=0.89 {\rm A/m},~d_1=d/2=700 {\rm nm},~d_2=875 {\rm nm},~G=1.4\times 10^5 {\rm T/m},
$$
where $d$ is the diameter of the ferromagnetic particle.
The frequencies $\Omega_n$ have been calculated for a silicon rectangular cantilever. (See Fig. 2.) 

\begin{figure}[t]
\centerline{\psfig{file=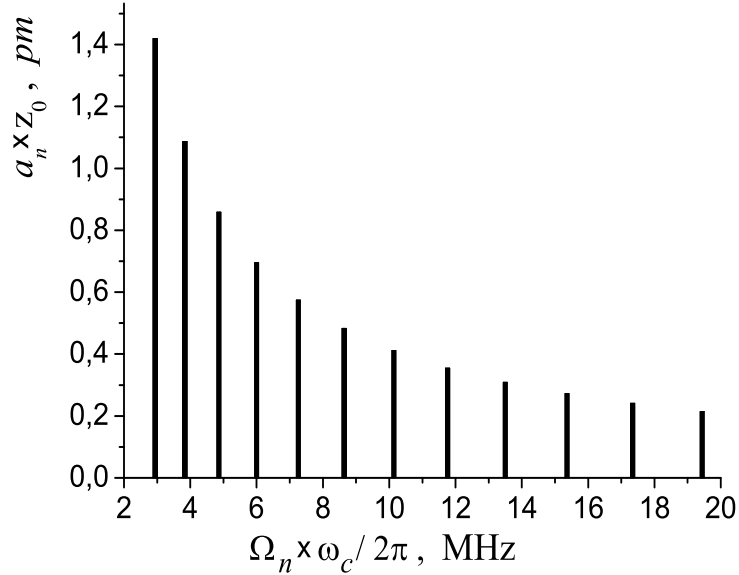,width=11cm,height=9cm,clip=}}
\vspace{4mm}
\caption{The thermal amplitudes of the high frequency cantilever modes for a silicon cantilever 
($190\mu m\times 3\mu m\times 850$nm) used in [2]. The left-most bar corresponds to the seventh harmonic. The Rabi frequency $\omega_R/2\pi=4.17$MHZ. The cantilever temperature is 80K.}
\label{fig:2}
\end{figure}

We calculate the relative shift of the cantilever period $\Delta T/\Delta T_0$ caused by the spins of the sample (where $\Delta T_0$ is the initial shift of the cantilever period). Initially all magnetic moments point approximately in the direction of the effective magnetic field. So, the initial shift $\Delta T_0$ is positive. The phases $\Psi_n$ in Eq. (5) were changed randomly in the interval $(0,2\pi)$. The characteristic time interval between the phase jumps was  taken as 10 Rabi periods $10\times(2\pi/\omega_R)$.   The number of high frequency modes taken into consideration was 22.

As an example, Fig. 3 demonstrates the connection between the regular and random MRFM signals, for a system of 50 magnetic moments uniformly distributed in the resonant slice. 

\begin{figure}[t]
\centerline{\psfig{file=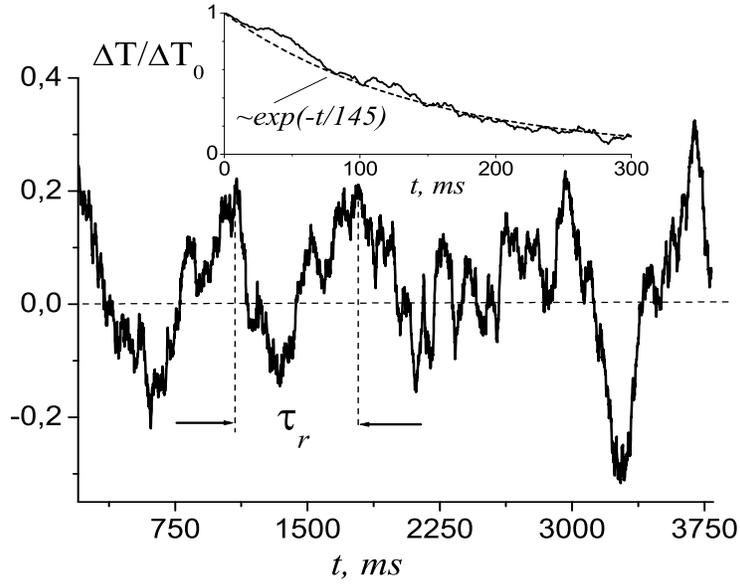,width=11cm,height=9cm,clip=}}
\vspace{4mm}
\caption{The random MRFM signal. The insert shows the regular MRFM signal. The temperature is 80K. The Rabi frequency $\omega_R/2\pi=4.17$MHz. The cantilever amplitude $z_0=28$nm.}
\label{fig:3}
\end{figure}
\begin{figure}[t]
\centerline{\psfig{file=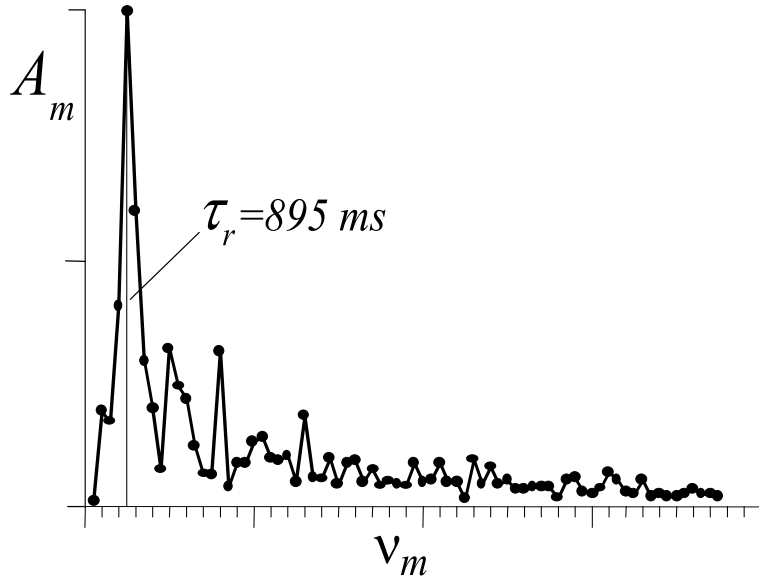,width=11cm,height=9cm,clip=}}
\vspace{4mm}
\caption{The Fourier spectrum of the random MRFM signal shown in Fig. 3.}
\label{fig:4}
\end{figure}
The decay time of the regular signal is $\tau_m=145$ ms. (The value of $\tau_m$ estimated using the formula (14) is 550 ms. Thus, our rough estimate provides the same order of magnitude as the computer simulations.) 
The MRFM random signal is about 30\%  of the maximum regular signal, and the characteristic time of the random signal $\tau_r=895$ms is not far from  $8\tau_m$. Fig. 4 shows the Fourier spectrum of the random signal (in arbitrary units): $\Delta T(t)/\Delta T_0=\sum A_m\cos(\nu_m t+\Phi_m)$. One can see a sharp peak near the value $1/\tau_r$. In fact, we find the value $\tau_r$ from the position of the peak in the Fourier spectrum. This position is a ``signature'' of the random MRFM signal. 

Fig. 5 shows the standard deviation of the random signal $\sigma_\xi=\sqrt{\overline{\xi-\bar\xi)^2}}$ (where $\xi(t)$ is the random function $\xi(t)=\Delta T(t)/\Delta T_0$) as a function of a number of magnetic moments N distributed in the resonant slice (for the same value of the average magnetization $M$). One can see that with increasing N the standard deviation $\sigma_\xi$ approaches the value 0.1.  This indicates that the random MRFM signal survives the transition to the continuous magnetization. Note that the position of the peak in the spectrum of the random signal (see Fig. 4) does not depend on N.

\begin{figure}[t]
\centerline{\psfig{file=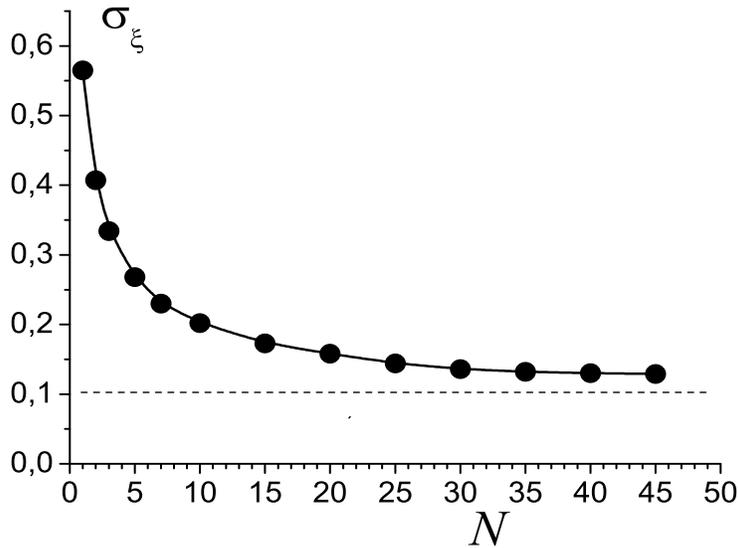,width=11cm,height=9cm,clip=}}
\vspace{4mm}
\caption{The standard deviation of the random MRFM signal as a function of the number N of magnetic moments (cells) in the resonant slice (at a fixed value of the average magnetization $M=0.89$A/m). All parameters are the same as in Fig. 3.}
\label{fig:5}
\end{figure}
\begin{figure}[t]
\centerline{\psfig{file=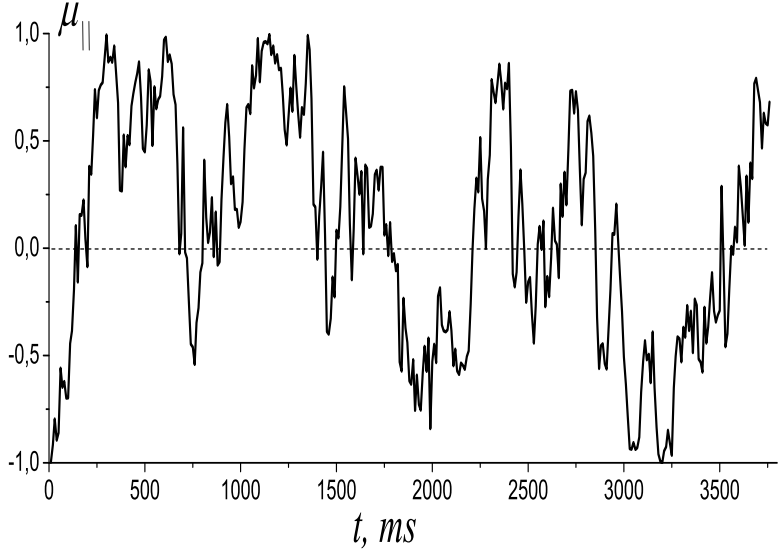,width=11cm,height=9cm,clip=}}
\vspace{4mm}
\caption{The component $\mu_{\|}(t)$ of an arbitrarily chosen magnetic moment along the effective magnetic field in the rotating frame. All parameters are the same as in Fig. 3.}
\label{fig:6}
\end{figure}

As an illustration to the spin dynamics, Fig. 6 shows the random change of the magnetic moment component along the effective field. (The magnetic moment was arbitrary chosen in the resonant slice.)  
One can see that the magnetic moment randomly moves between the direction of the effective field and the opposite direction. The characteristic time of this motion $\tau_r$ is the same as for the MRFM signal shown in Fig. 3.

As an extreme example, we have considered the case when the cantilever amplitude is only 60 pm. In this case, the initial magnetic moment of the resonant slice for our value of the magnetization $M=0.89$ A/m, is approximately one Bohr magneton. 
Figs. 7 and 8 show the regular and random signals for the number of  magnetic moments (cells in the resonant slice) N=200 and N=1000. One can see that the relaxation time of the regular signal $\tau_m\approx 5.5$ ms is the same for both values of N. 

\begin{figure}[t]
\centerline{\psfig{file=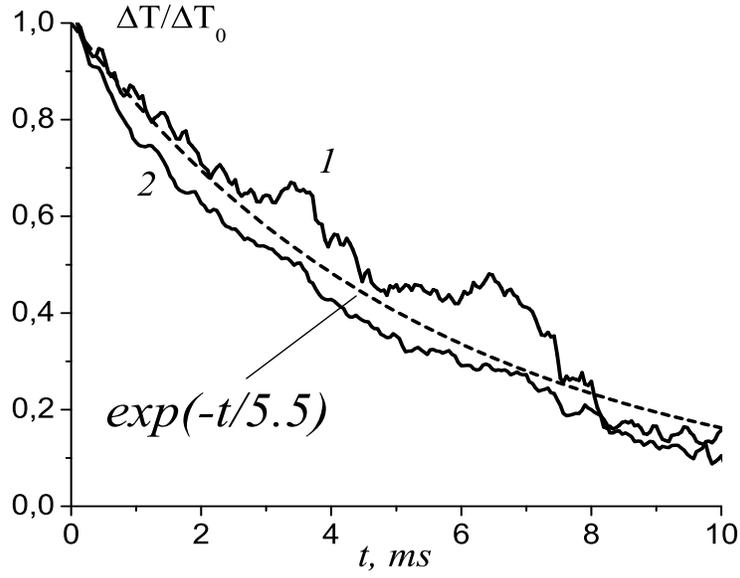,width=11cm,height=9cm,clip=}}
\vspace{4mm}
\caption{A regular MRFM signal for an extremely small amplitude of the cantilever vibrations, $z_0=60$pm. The temperature of the cantilever is 0.57K. The Rabi frequency $\omega_R/2\pi=8.34$MHz. Curve 1: N=200; curve 2: N=1000.}
\label{fig:7}
\end{figure}
\begin{figure}[t]
\centerline{\psfig{file=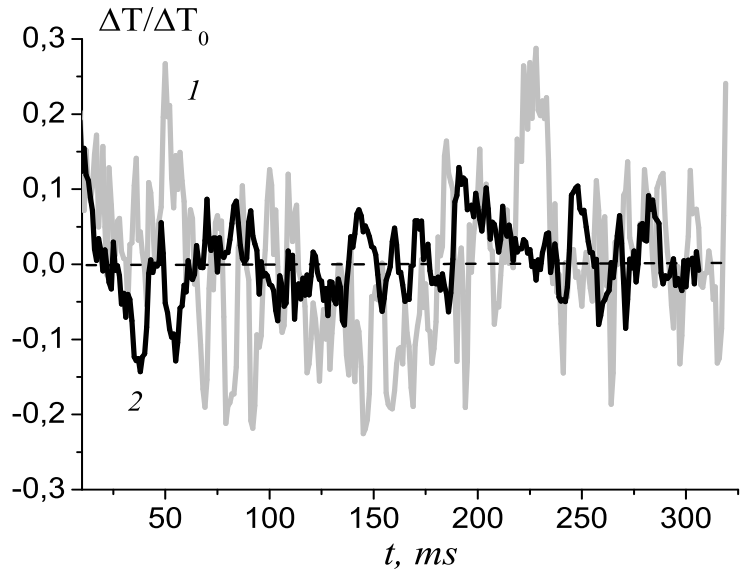,width=11cm,height=9cm,clip=}}
\vspace{4mm}
\caption{The random MRFM signal, for the same parameters as in Fig. 7. Curve 1: N=200; curve 2: N=1000.}
\label{fig:8}
\end{figure}

The value of the standard deviation $\sigma_\xi$ for the random signal drops from 0.105 to 0.052, when N increases from N=200 to N=1000. The search for a ``plateau'' in the dependence, $\sigma_\xi(N)$, requires a greater number of the  magnetic moments (cells) and, correspondingly, much greater computational time. The spectrum of the random MRFM signal for both values of N has its maximum at a frequency of approximately 17.5 Hz, which  corresponds to the characteristic time $\tau_r=57$ ms. This is not far from the estimated value $\tau_r\sim 8\tau_m=44$ ms. Note, that for the extremely small amplitude of the cantilever vibrations considered here, instead of adiabatic reversals we have small adiabatic oscillations of the effective magnetic field near the $x$-axis in the rotating frame. 

\section*{Conclusion}

In this paper we study the random non-dissipative MRFM signal caused by the high frequency thermal vibrations of the cantilever. For a single spin detection, 
this random signal is expected to have the same amplitude as the initial regular signal. We show that for a relatively large number of spins in the resonance slice the random non-dissipative signal occurs after the relatively short regular signal. Our computer simulations indicate that even for a very large number of spins, when one can use the approximation of a continuous magnetization, a long lasting random signal can be observed, until the intrinsic mechanisms of spin relaxation destroy the local magnetization in the resonant slice. The random non-dissipative MRFM signal can be used for the spin detection, including a single spin detection. 

\section*{Acknowledgments}

We thank D. Rugar for his continuing support and helpful discussions. This work  was supported by the Department of Energy under the contract W-7405-ENG-36 and DOE Office of Basic Energy Sciences, by the National Security Agency (NSA), by the Advanced Research and Development Activity (ARDA), and by the DARPA Program MOSAIC. 

{}
\end{document}